\documentstyle[epsfig,twocolumn,aps]{revtex}

\begin{document}

\title{Different Gd$^{3+}$ sites in CaB$_{6}$: an ESR study}
\author{R.R. Urbano,$^{1}$ C. Rettori,$^{1}$ G.E. Barberis,$^{1}$ M. Torelli,$^{2}$ A. Bianchi,$^{2}$ Z.
Fisk,$^{2}$ P.G. Pagliuso,$^{3}$ A. Malinowski,$^{3}$ M.F.
Hundley,$^{3}$ J.L. Sarrao$^{3}$ and S.B. Oseroff,$^{4}$ }
\address{$^{1}$Instituto de F\'{\i}sica ``Gleb Wataghin'', UNICAMP,13083-970,
Campinas-SP, Brazil.\\ $^{2}$National High Magnetic Field
Laboratory, Florida State University,\\ Tallahassee, FL 32306,
U.S.A. \\ $^{3}$ Los Alamos National Laboratory, Los Alamos, New
Mexico 87545, U.S.A. \\ $^{4}$San Diego State University, San
Diego, CA, 92182, U.S.A.}

\maketitle

\begin{abstract}
The local environment of Gd$^{3+}$ ($4f^{7}$, $%
S=7/2$) ions in single crystals of Ca$_{1-x}$Gd$_{x}$B$_{6}$ ($0.0001$ $%
\lesssim x\lesssim 0.01$) is studied by means of Electron Spin
Resonance (ESR). The spectra for low concentration samples
($x\lesssim 0.001$) show a split spectrum due to cubic crystal
field effects(CFE). The lineshape of each fine structure line is
lorentzian, indicating an insulating environment for the Gd$^{3+}$
ions. For higher concentrations ($0.003$ $\lesssim x\lesssim
0.01$), the spectra show a single resonance ($g=1.992(4),$ $\Delta
H_{1/2}\approx 30-60$ Oe) with no CFE and dysonian lineshape
indicating metallic environment for the Gd$^{3+}$ ions. For
intermediate concentrations, a coexistence of spectra
corresponding to insulating and metallic regions is observed.
Most of the measured samples show the weak ferromagnetism (WF) as
reported for Ca$_{1-x}$La$_{x}$B$_{6}$ ($x\approx 0.005$), but,
surprisingly, this WF has no effect in our ESR spectra either for
metallic or insulating environments. This result suggests that
the ferromagnetism in these systems might be isolated in clusters
(defect-rich regions) and its relationship with metallicity is
nontrivial.
\end{abstract}

\pacs{71.10.Ca,71.35.-y,75.10.Lp}

\narrowtext

The weak ferromagnetism (WF) and Curie temperature of roughly
$600$ K recently reported by Young {\it et al}., \cite{Young} in
lightly-La doped calcium hexaborides (Ca$_{1-x}$La$_{x}$B$_{6}$
($x\cong 0.005$) has fueled interest from theorists and
experimentalists. It has, for example, revived interest in the
physics of the so called {\it ''excitonic insulators'',}
vigorously discussed in the 60's and 70's. \cite{Halperin}
However, recently, controversial experimental results have been
reported creating extensive discussion. For example, Vonlanthen
et al.\cite{Vonlanthen} reported that a CaB$_{6}$ sample close to
stoichiometry does not show WF but instead one deficient in Ca is
magnetic. Moriwaka et al.\cite{Moriwaka} argued that FM is
induced by Ca vacancies in CaB$_{6}$ and speculated that a
lowering of the crystal symmetry may play a role in its origin.
Monnier et al. \cite{Monnier} claimed that a neutral B$_{6}$
vacancy would be more effective to induce FM. Giann\`{o} et.
al.\cite{Giannio} suggested that the defects are predominantly
donors, providing partly itinerant and partly localized
electrons. Terashima et al. reported that the data in
Ca$_{0.995}$La$_{0.005}$B$_{6}$ is sample
dependent.\cite{Terashima} Kunii\cite{Kunii} observed a highly
anisotropic ferromagnetic resonance(FMR) line in
Ca$_{0.995}$La$_{0.005}$B$_{6}$ and concluded that the spins are
confined within a surface layer of $\sim$ 1.5 um thick, which
leads to a magnet moment several orders of magnitude larger than
that obtained for the bulk. Denlinger et al.\cite{Denlinger}
measured photoemission spectroscopy and found a large difference
between the electronic structure of the surface and that expected
for the bulk.  All these results have raised many questions about
the remarkable properties of these hexaborides compounds, such
as: i) Is the WF in the bulk or just in the surface of the
material? ii) What kind of defects are present in these
compounds, and how do they dope these materials? iii) Is there a
FMR in Ca$_{0.995}$La$_{0.005}$B$_{6}$? iv) Is there is a
reduction in the cubic symmetry of the crystal or change in the
EPR spectra at the ordering temperature? and v) What kind of
correlation exist between magnetic and transport properties in
these hexaborides?

Despite the experimental controversies, recent ambitious
theoretical communications using different approaches have tried
to explain the WF in these materials.
\cite{Zhitomirsky,Leon,Victor,Jarlborg,Ceperley,Tromp} In
particular, Tromp et al. suggest that CaB$_{6}$ is a
semiconductor with a band gap of 0.8 eV and that the magnetism in
the La doped hexaborides occurs on the metallic side of a Mott
transition in the La-induced impurity band.\cite{Tromp} In
comparison, earlier models rely on electronic structure
calculations, which claimed that CaB$_{6}$ is a semimetal with a
small overlap of the conduction and valence bands at the X point
of the Brillouin
zone.\cite{Zhitomirsky,Leon,Victor,Jarlborg,Ceperley} However, a
complete and conclusive modeling of this phenomenon that includes
all of the experimental results is still missing.

In this letter we present ESR, magnetization and resistivity results in Ca$_{1-x}$Gd$_{x}$B$_{6}$ ($%
0.0001\lesssim x\lesssim 0.01$). ESR is a powerful local
experimental technique that can used to probe the microscopic
properties of these materials. Our main findings are: $i$) an
unusual CFE is observed in the ESR spectra of diluted Gd$^{3+}$ in CaB$%
_{6}$; coexistence of Gd$^{3+}$ sites {\it with} and {\it without}
crystal field (CF) splitting is found, $ii$) the ESR spectra of
Gd$^{3+}$ are not affected by the presence of WF in these systems,
$iii$) We found no evidence for a reduction of the cubic symmetry
in these compounds, and $iv$) the WF in these materials is a bulk
property which is strongly sample and doping dependent with
nontrivial correlation with metallicity. As we will discuss, these
results serve to substantially constrain the range of models
applicable to the WF in these materials.

More than fifty single crystalline Ca
excess/deficient/stoichiometric samples of
Ca$_{1-x}$R$_{x}$B$_{6}$ ($0.0001\lesssim x\lesssim 0.01$, R =
La, Gd, Er) were grown as described in ref. \cite{Young}. The
structure and phase purity were checked by x-ray powder
diffraction and the orientation was determined by Laue x-ray
diffraction. The ESR experiments were carried out in a Bruker
X-band spectrometer, using a TE$_{102}$ room-$T$ cavity and a
helium gas flux ($4.2$ K $-$ $300$ K) temperature controller. The
magnetization measurements between $2$ K and $800$ K were made in
a Quantum Design SQUID $dc$-magnetometer. Electrical resistivity
was measured using a low-frequency $ac$-resistance bridge and
4-contact configuration. The concentration for all the samples
reported in this work was deduced from Curie-Weiss fits of the
low-$T$ susceptibility data.

Results shown here span the full range of properties observed
among the more than fifty samples studied. Figures 1a, 1b, and 1c
show the ESR, $M_{dc}(T)$, and $M_{dc}(H)$ data for three of our
Ca$_{1-x}$Gd$_{x}$B$_{6}$ samples ($x=0.0003,0.003,0.01$).
Electrical resistivity, $\rho$($T$), data for the $x=0.0003$ and
$x=0.003$ are also shown in Fig.1. All these samples show a WF
component and $T_{C}$ between $600$ K and $800$ K. The data are
qualitatively similar to that of Ca$_{0.995}$La$_{0.005}$B$_{6}$
reported by \ Young {\it et.al}. \cite{Young} However, the WF
component (saturation magnetic moment) for the $x=0.0003$ and
$x=0.003$ samples is estimated to be $\sim 3$ and $\sim 0.16$
$\mu _{B}$/Gd, respectively, which are much larger than that
reported in ref. 1 and there is no clear systematic between $x$
and the WF component.

For X-band measurements on insulators, the Gd$^{3+}$ ions in a
cubic-site symmetry usually show a characteristic CF split ESR
spectrum with fine structure of seven lorentzian lines of well
defined relative intensities due to their transition probabilities
and population Boltzmann factor. \cite{Bleaney} The different
Gd$^{3+}$ spectra of Fig. 1 show that there are two different
Gd$^{3+}$ sites. For samples of low concentration ($x\lesssim
0.0003$, Fig. 1a) the spectra show the typical fine structure due
to cubic CF of insulators. The high concentration samples
($0.003\lesssim x\lesssim 0.01$, Fig. 1c) show a single resonance
with a dysonian lineshape ($A/B$ $\approx 2.2$) characteristic of
skin depth smaller than the size of the sample. \cite{Feher} No
CFE were observed in this resonance. Absorption/dispersion
lineshape analysis \cite{Pake} of this
resonance between $4.2$ and $50$ K yields a $T$-independent $g$-value ($%
g\approx $ $1.990(6)$) and linewidth ($\Delta H_{1/2}\approx 80$
Oe). For samples of intermediate concentration ($0.0003\lesssim
x\lesssim 0.003$, Fig. 1b) the ESR spectra show the coexistence
of both Gd$^{3+}$ sites. In addition, $\rho$($T$) data  for
$x=0.0003$ and $x=0.003$ taken on different crystals from the same
batches (with the same measured WF component) show metallic-like
behavior and comparable absolute values for both $x=0.0003$ and
$x=0.003$ samples.

Figures 2a and 2b show the ESR spectra for a set of samples
similar to those of Fig. 1. Although the ESR spectra show similar
features ($g$-values, linewidths, and two Gd$^{3+}$ sites) all of
them presented smaller saturation magnetic moment ($\lesssim $
$0.5$ emu/mole f.u.) than those of Fig. 1. Also, we found that
some of the undoped Ca excess/deficient/stoichiometric samples
presented an even smaller, but measurable, saturation magnetic
moment ($\lesssim $ $0.1$ emu/mole f.u.). In this case, samples
are found consistently to be insulator-like by ESR and
resistivity.(see data for $x=0.0001$ in Fig.2a) Hence, it is clear
from the results of Figs. 1 and 2 that the Gd-doping contributes
to the WF in these systems, but it is also true that the WF is
strongly sample dependent.

In some of the as grown crystals we also observed an extra
isotropic resonance at $g=$ $2.0028(6)$ with a lorentzian
lineshape and linewidth of $\Delta H_{pp}\approx 7$ Oe. This
resonance could be eliminated by gently etching and/or polishing
the crystal surfaces (see Fig. 2a for $x=0.0001$). We associate
this resonance with an unidentified ESR active center on the
surface of the crystals. Surface etching and/or polishing $did$
$not$ $affect$ the ESR spectra or $M_{dc}(T,H)$ data shown in
Figs. 1 and 2. Fig. 2b shows the ESR spectra for one of the lowest
Gd concentration samples ($x\lesssim 0.00005$). This spectrum
shows, besides the small
resonances of Gd$^{3+}$, the typical spectrum due to natural impurities of Eu%
$^{2+}$ in cubic symmetry with fine and hyperfine structure
($4f^{7}$, $S=7/2 $, $^{151}$Eu$^{2+}$ and $^{153}$Eu$^{2+}$
isotopes, $I=5/2$). \cite{Kunii2}

Figure 3a presents the angular dependence in the (110) plane of
the ESR spectra of Figure 3b. This angular dependence corresponds
to that expected for Gd$^{3+}$ ions, ($4f^{7}$, $S=7/2$), with
cubic site symmetry. \cite {Bleaney} The solid lines are fits
obtained from the Spin Hamiltonian, ${\cal H}=g\beta
HS+b_{4}O_{4}+b_{6}O_{6} $, to the data assuming cubic CF and an
isotropic $g$-factor.\cite{Bleaney} The different intensity
between the transitions $-5/2\longleftrightarrow -7/2$ and
$7/2\longleftrightarrow 5/2$ at low-$T$ leads to a negative value
for $b_{4}$ (see Fig. 3b). The Gd$^{3+}$ fitting parameters are:
$g =1.992(4)$, $b_{4}$ = - 13.3(5) Oe, and $b_{6}$ $<$ 1(1) Oe.
The simulation of the Gd$^{3+}$ spectrum for $H//[001]$ is shown
in Fig. 3b. It is clear from this spectrum that two Gd$^{3+}$
sites are coexisting in this material for
intermediate Gd concentrations, one {\it with} CF splitting and another {\it %
with no} CF splitting. The Spin Hamiltonian parameters were
concentration independent, within the accuracy of the
measurements.

The results obtained in the 0.5\% co-doped La and Ca excess/deficient Ca$%
_{1-x}$(Gd,Er)$_{x}$B$_{6}$ samples showed the general behavior
and features described in Figs. 1-3. For the sake of simplicity,
these data are not included in the figures. We want to mention
that ESR in Ca$_{1-x}$La$_{x}$B$_{6}$ for $x=0$ and $x\cong
0.005$, showing saturation magnetic moment of $\sim 0.1$ and
$\sim 1-2$ emu/mole f.u., respectively, did not show any traces of
other-R natural impurities.

The observation of a coexistence of different Gd$^{3+}$ sites,
with ESR spectra of different lineshape and structure, suggest
that there are distribution of doping levels along these samples.
The presence of Gd$^{3+}$ resonances with lorentzian (insulator)
and dysonian (metallic) lineshapes and different CFE lead us to
conclude about the coexistence of low and highly doped regions in
the samples. Also, crystals from the same batch (with similar WF)
were found to be metallic-like by $\rho$($T$) and insulator-like
by ESR (see $x=0.0003$ in Fig. 1) confirming strong doping
inhomogeneity for these range of concentration($ 0.0003\lesssim
x\lesssim 0.003$). An explanation for the observed sample
dependency and doping inhomogeneity is probably related to the
presence of defects in the sample. Defects can either dope the
sample with carriers or serve as deep local traps capturing local
moments on the order of 1 $\mu_{B}$. Vacancies of Ca or neutral
B$_{6}$ have been suggested as the origin for these
effects.\cite{Moriwaka,Monnier,Giannio}. Whatever their origin,
they certainly can be strongly growth- and sample-dependent.
Therefore, for low/intermediate concentrated samples defects
levels can play an important role causing significant spatial
doping inhomogeneity.
Consistently, for more concentrated samples, $%
0.003\lesssim x\lesssim 0.01$ (where the effective doping is
probably more spatially homogeneous), the ESR results (single
dysonian resonance) were only slightly sample dependent and always
consistent with metallic $\rho$($T$) data.

On the other hand, the WF\ component was still strongly sample
dependent (see Fig. 1-2) even for samples where both ESR and
resistivity have shown metallicity. Samples with similar
resistivity behavior and absolutes values ($x=0.0003$, $x=0.003$
and $x=0.016$) have revealed very different (up to one order of
magnitude) WF component(see Fig. 1 - 2). Similarly, samples with
different transport properties ($x=0.0001$ and $x=0.0016$)
revealed similar WF component. These results show the
relationship between metallicity and magnetism is not trivial. As
suggested by thermopower data,\cite{Giannio} defects can randomly
create carriers or trapped local moments. Furthermore, for both
metallic and insulator Gd$^{3+}$ sites, the ESR is completely
unaffected by the WF component suggesting that FM develops in
isolated clusters (probably defect-rich regions in the sample)
which could make the exchange interaction between the Gd$^{3+}$
ions and the WF moments negligible. On the other hand, the
Gd$^{3+}$ ESR data is strongly sensitive to the increase of
carriers in its environment, showing lineshape changes and an
intriguing collapse of the fine structure (CEF splitting).
Therefore a metallic $island$ in the sample can exist without
necessarily display a magnetic moment creating a non-obvious
correlation between magnetic and transport properties.

The absence of Gd$^{3+}$ $g$-shift (Knight shift), thermal
broadening of the linewidth (Korringa rate), and long estimates for ${\cal T}%
_{1}$ (spin-lattice relaxation time)\cite{Standley} in these
metallic regions suggest that the exchange interaction between
the rare-earth localized magnetic moments and conduction
electrons may be negligible in these regions.
\cite{Rettori,Taylor} Therefore, the absence of CFE in the
Gd$^{3+}$ spectra of these metallic regions is unlikely to be due
to CF exchange narrowing mechanisms. \cite{Plefka,Barnes,Urban}
Also, the relatively low Gd concentration used in this work
($x\lesssim 0.01$) and the observed $T$-independent $g$-value and
linewidth suggests that these metallic region cannot be
associated with clusters of Gd$^{3+}$ ions. As such, the absence
of CFE in the Gd$^{3+}$ ESR spectra is an unexpected result that,
to the best of our knowledge, has only been previously reported
in one system, YBiPt, which perhaps not coincidentally is a small
gap semiconductor.\cite{Pagliuso} Since the Er$^{3+}$ resonance
remains a $\Gamma _{6}$ crystal field ground state even in these
metallic regions (not shown), the absence of CFE in the \
Gd$^{3+}$ ESR spectra of these materials should be intrinsically
associated with the basic mechanisms that lead to the observation
of fine structure in a $S$-state ground state.\cite{Bleaney}
\cite{Baberschke} It is clear from our data that these mechanisms
must to be reviewed.

An alternative speculation about our findings would come from the
comparison of the phase separation predicted in refs. 11 and 12
with the insulating/metallic regions reported here, suggesting
that the samples of Fig. 1 and 2 present SDW (triplet) and CDW
(singlet) excitonic condensate ground states, respectively.\cite
{Victor} However, concerns about the excitonic picture arose
after WF was also in defect doped hexaborides.\cite{Moriwaka}This
result would seem to rule out the idea that the $d$-electrons of
La atoms play a fundamental role to the magnetism in this
compounds.

In this work we employed ESR, one of the most powerful local
experimental technique, to investigate the microscopic properties of the Ca$%
_{1-x}$Gd$_{x}$B$_{6}$ ($0.0001\lesssim x\lesssim 0.01$) system.
The systematic ESR, $M_{dc}(T)$, $M_{dc}(H)$ and $\rho$($T$) study
of more than fifty samples reveals that the relatively strong WF
observed in these materials is a bulk property resulting from a
subtle combination of R-doping and sample preparation methods. In
agreement with previous work,\cite{Moriwaka} we find that some of
the R-undoped samples prepared by different methods also showed
the presence of WF, presumably due to defect self-doping. The ESR
experiments show that for $x\lesssim 0.003$ the samples are
inhomogeneous presenting a coexistence of insulating and metallic
regions. We believe this happens when doping is dominated by
defects. Further, ESR, $M_{dc}(T,H)$ and $\rho$($T$) data are
extremely sample dependent, and as far as the doping is concerned
Gd ions behave like La ions. Within the accuracy of our experiments ($%
\sim 10\%$ of the measured linewidths, $\sim 10$ Oe), we found no
evidence for a WF-induced $g$-shift in the Gd$^{3+}$ ESR spectra.
Moreover, although the resonances in the metallic regions are
broader than in the insulating regions, no correlation between the
ESR linewidth and WF component was found even for high
concentrated metallic samples, which suggest that WF in these
compounds might be associated with coupled local moments isolated
in clusters. Nevertheless, the striking results showing
coexistence of microscopic regions {\it insulator/with} and {\it
metallic/without} CFE in the Gd$^{3+}$ ESR spectra suggest that
the interplay between CF, self-doping, R-doping, conductivity and
magnetism is very subtle, and it is at the moment far from
understood in these systems. In particular, the origin and nature
of the self-doping defects need to be clarified and included in
the models.

{\it \bigskip }

This work was supported by FAPESP, CNPq, NFS and US DOE.

\narrowtext

\begin{figure}[tbp]
\centerline{\includegraphics[scale=0.35]{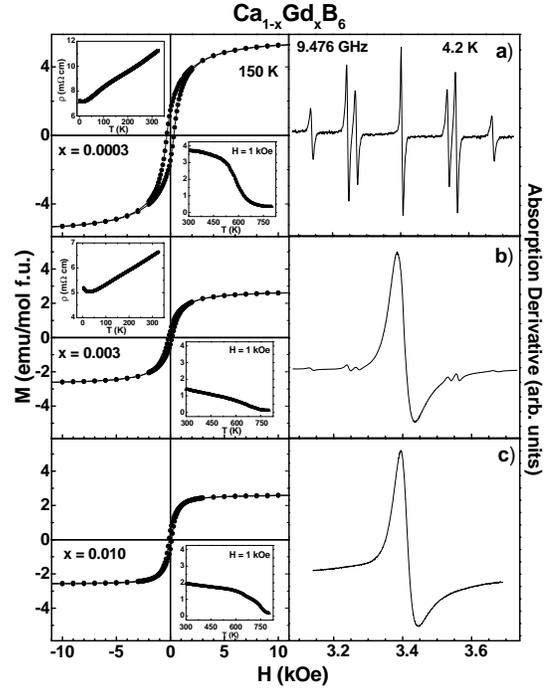}} \caption{ESR,
$M_{dc}(T)$, and $M_{dc}(H)$ data for three samples of
Ca$_{1-x}$Gd$_{x}$B$_{6}$: a) $x=0.0003$, b) $x=0.003$, and c)
$x=0.01$. $\rho$($T$) data for the $x=0.0003$ and $x=0.003$ are
also shown} \label{Figure 1}
\end{figure}

\begin{figure}[tbp]
\centerline{\includegraphics[scale=0.35]{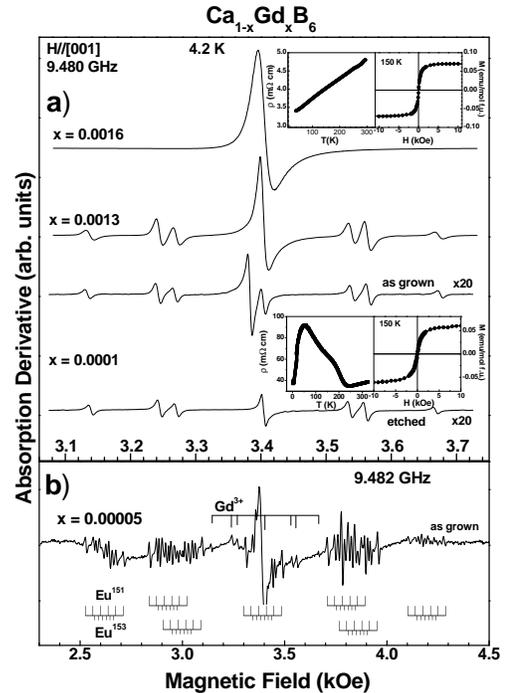}}
\caption{Gd$^{3+}$ ESR spectra for four samples of
Ca$_{1-x}$Gd$_{x}$B$_{6}$%
: a) $x=0.0016$, $x=0.0013$, $x=0.0001$ (as grown and etched) and
$\rho$(T) and $M_{dc}(H)$ data for the $x=0.0016$ and $x=0.0001$.
b) $x=0.00005$ (as grown). This sample shows the ESR spectrum of
natural impurities of Eu$^{2+}$.} \label{Figure 2}
\end{figure}

\begin{figure}[tbp]
\centerline{\includegraphics[scale=0.30]{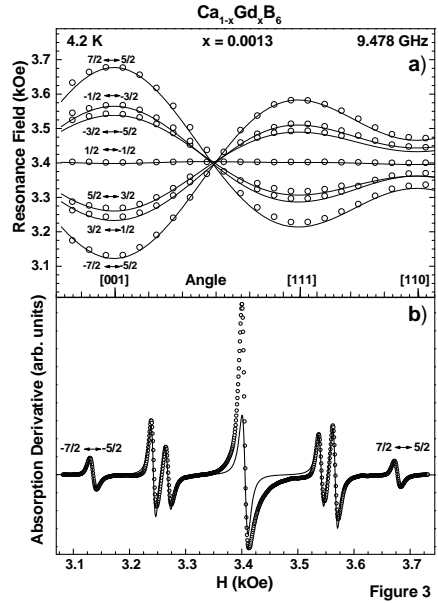}} \caption{ESR
data for Ca$_{1-x}$Gd$_{x}$B$_{6}$ ($x=0.0013$): a) angular
dependence of the field for resonance in the ($110$) plane and, b)
Experimental(open circles) and simulated (solid line) ESR spectrum of Gd$^{3+}$ for $%
H//[001]$.} \label{Figure 3}
\end{figure}

\end{document}